# Atomic structure, electronic structure and optical absorption of inorganic perovskite compounds $Cs_2SnI_{6-n}X_n$ (X=F, Cl, Br; n= 0~6): A first-principles study

Wang Xuan[1], Tang Yehua[1], Nairui Xiao, Wang Ke-Fan[*]

Henan Key Laboratory of Photovoltaic Materials, Henan University, Kaifeng 475004, P. R. China

**ABSTRACT**: As a possible alternative to organic-inorganic hybrid perovskite halide, inorganic $Cs_2SnI_6$ has drawn more and more research attention recently. In order to find more $Cs_2SnI_6$ derivatives as the potential solar cell absorber materials, $I^-$ ions in $Cs_2SnI_6$ are replaced by other halogen ions and forms the $Cs_2SnI_{6-n}X_n$ (X=F, Cl, Br; n=1~6) compounds, whose atomic structures, electronic structures and optical absorption are investigated by first principles calculation. When the alloying level *n* increases, the mean lattice constants, the weighted Sn-X and Cs-X bond lengths all decreases linearly; the bond length of each Sn-X diminishes slightly inside the octahedral structure; $E_g$ of $Cs_2SnI_{6-n}X_n$ increases nonlinearly. Eleven $Cs_2SnI_{6-n}X_n$ compounds have an $E_g$ between 1.0 eV and 2.0 eV and so can be potentially used as the light absorption layer of solar cells. Their partial DOS demonstrate that as the alloying level *n* increases, I 5p orbital in VBM and CBM is gradually substituted by Br 4p, or Cl 3p, or F 2p orbital. The eleven $Cs_2SnI_{6-n}X_n$ alloys all have a *direct* bandgap although the lattice distortion induced by the alloyed $X^-$ ion.

[*] Corresponding author. E-mail addresses: kfwang@henu.edu.cn (Prof. Ke-Fan Wang).

[1] These authors contribute equally to this work.

## 1. Introduction

In the last twelve years, organic-inorganic hybrid perovskite halides have become the star light-absorbing material for the next generation solar cells, primarily due to the quick leaping of power conversion efficiency (PCE), easy and cheap fabrication technique for its solar cells. Its PCE has been enhanced from 3.8% in the year 2009 to 25.5% at present [1, 2]. However, the instability of organic group and the usage of toxic element Pb in such hybrid perovskites have prevented their high-efficiency solar cells from widely using. In order to solve the two issues, stable and nontoxic inorganic perovskite halides (including but not limited to $CsSnI_3$ [3, 4], $CsBX_3$(B=Sn, Ge; X=I, Br, Cl) [5], $A_2BX_6$ (A = K, Rb, Cs; B = Si, Ge, Sn, Pb, Ni, Pd, Pt, Se and Te; X = Cl, Br, I;) [6], $Cs_2TiI_xBr_{6-x}$ [7]) has been suggested to replace their organic-inorganic counterpart recently.

$Cs_2SnI_6$, a vacancy ordered double inorganic perovskite without the toxic Pb element, owns a highest oxidation state of tin ($Sn^{4+}$) and thus an intrinsic resistance to oxidization and hydrolysis. It can exist stably for one hour at 150 ºC in air [8] and so a much better thermal stability than that of the organic-inorganic hybride perovskite, such as $MAPbI_3$ or $MA_{0.7}FA_{0.3}Pb(I_{0.9}Br_{0.1})_3$ [9]. It is widely accepted that $Cs_2SnI_6$ has a bandgap of ~1.3 eV [10-14], which will promise a theoretical Shockley-Queisser efficiency of ~30% for its solar cells [15]. At present, however, only a PCE of 0.96% has been reported for pure $Cs_2SnI_6$ perovskite solar cells [16]. When $Cs_2SnI_6$ is partially alloyed by $Br^-$ ion to form $Cs_2SnI_4Br_2$, its solar cell owns a larger PCE of 2.1% and its air stability was also improved [17]. Consequently, we believe that the substituting of $I^-$ ions with other halide ions may be an effective method to improve its PCE and stability.

About the effect of other halogen ions on the physical properties of $Cs_2SnI_6$, there are a few literatures published recently, either about the effect of the $Br^-$ ion [14, 18-19] or of the $Cl^-$ ion [12, 20]. These studies focus on the experimental investigations of the $Cs_2SnI_{6-x}Br_x$ or the Cl-enriched $Cs_2SnI_xCl_{6-x}$ compounds. Till now, a comprehensive theoretical investigation about the effects of halogen ions on the $Cs_2SnI_6$ is still unavailable. In this paper, we will replace the $I^-$ ions in $Cs_2SnI_6$ by $F^-$, $Cl^-$ and $Br^-$ ions to form the $Cs_2SnI_{6-n}X_n$ (X=F, Cl, Br; n=1~6) compounds, and then investigate their atomic structures, electronic structures and optical absorption via first-principles calculations. Before the expensive and time-consuming experimental investigations, the theoretical exploring of the physical properties of new materials by first-principles calculations is usually convenient and cheap. Moreover, the theoretical calculations are also helpful to understand the existing experimental results from the atomic or electronic viewpoint.

## 2. Computational methods and details

We used the primitive cell of $Cs_2SnI_6$ as the origin structure and then its six $I^-$ ions were substituted one by one by other halogen ions ($F^-$, $Cl^-$, or $Br^-$) to simulate the different alloying levels. All of the first-principles calculations were carried out by density-functional theory (DFT) [21,22] method implemented plane-wave-based Vienna ab initio simulation package (VASP) [23,24]. For the structure relaxations and static energy calculations, we used the generalized gradient approximation (GGA) through the Perdew, Burke, and Ernzerhof (PBE) [25] functional to consider the exchange-correlation potential. A 5×5×5 Γ-centered Monkhorst-Pack grid of reciprocal lattice points was used to sample the irreducible Brillouin zone (BZ). The cut-off energy for the plane-wave basis is set as 550 eV for F-contained perovskites, and 400 eV for other halide perovskites. The ion coordinates are relaxed freely with an energy convergence threshold of $10^{-6}$ eV per atom and a residual force tolerance of 0.001 eV/Å on each atom. The lattice constant of $Cs_2SnI_6$ determined with these parameters is 12.065 Å, which is 3.74% larger than the experimental value of 11.63~11.65 Å [11, 14, 26-27], due to the overestimated lattice constants by PBE potential [28]. We also tested the PBEsol functional and calculated the lattice constant of $Cs_2SnI_6$ to be 11.558 Å that is very close to the experimental value of 11.63 Å, but it gives a much smaller bandgap energy of 0.6 eV than that by using PBE functional (1.05 eV, both plus HSE06 functional) or the experimental bandgap of ~1.3 eV. So we choose the PBE functional to consider the exchange-correlation potential in the end.

For the band structure calculations, the PBE and HSE06 functions were used respectively. In HSE06, the exact exchange is separated into a long-range part, which is essentially described by PBE, and a short range part, which is mixed with the Hartree-Fock (HF) and PBE exchanges. In our calculations, the screening parameter of μ=0.2 is used, and the chosen mixing coefficient is 25%. The $Cs_2SnI_6$ band gap obtained by HSE06 is 1.05 eV, which is a little smaller than the experiment value of ~1.3 eV, due to the used theoretical lattice constant of 12.065 Å.

The optical absorption is obtained from the dielectric function. The imaginary part of dielectric function is obtained by summing over the independent transitions between Kohn-Sham states neglecting local field effects [29], the real part was obtained from the imaginary part making use of the Kramers-Kronig relation with a small complex shift η = 0.1. Approximately 100 empty bands and a Γ-5×5×5 Monkhorst-Pack sampling of the Brillouin zone are needed for the convergence of the optical properties.

## 3. Results and discussion

### 3.1. Atomic structures of $Cs_2SnI_{6-n}X_n$ alloy

Fig. 1(a) shows the atomic structure of the $Cs_2SnI_6$ primitive cell (PC) before the alloying with other halogen ions. Six $I^-$ ions are substituted one by one by the halogen ions ($F^-$, or $Cl^-$, or $Br^-$), and then the alloyed PCs are relaxed completely. Then we find the symmetry of the relaxed PCs with a tolerance of 0.1 Å and impose the symmetry to the PC structures. Finally, we obtain the conventional cells after some trimming, for examples, as shown in Fig. (b)-(h) for $Cs_2SnI_{6-n}Br_n$. Fig. 1(b) shows the conventional cell of $Cs_2SnI_6$, in which $Sn^{4+}$ ion bonds with six $I^-$ ions in an octahedra structure while $Cs^+$ ion bonds with twelve $I^-$ ions in the cuboctahedral interstices. Fig. 1(c)-1(h) show the relaxed $Cs_2SnI_{6-n}Br_n$ conventional cells. Crystallographic information files (CIF) of the relaxed atomic structures of $Cs_2SnI_{6-n}X_n$ are supplied in Supplemental Information.

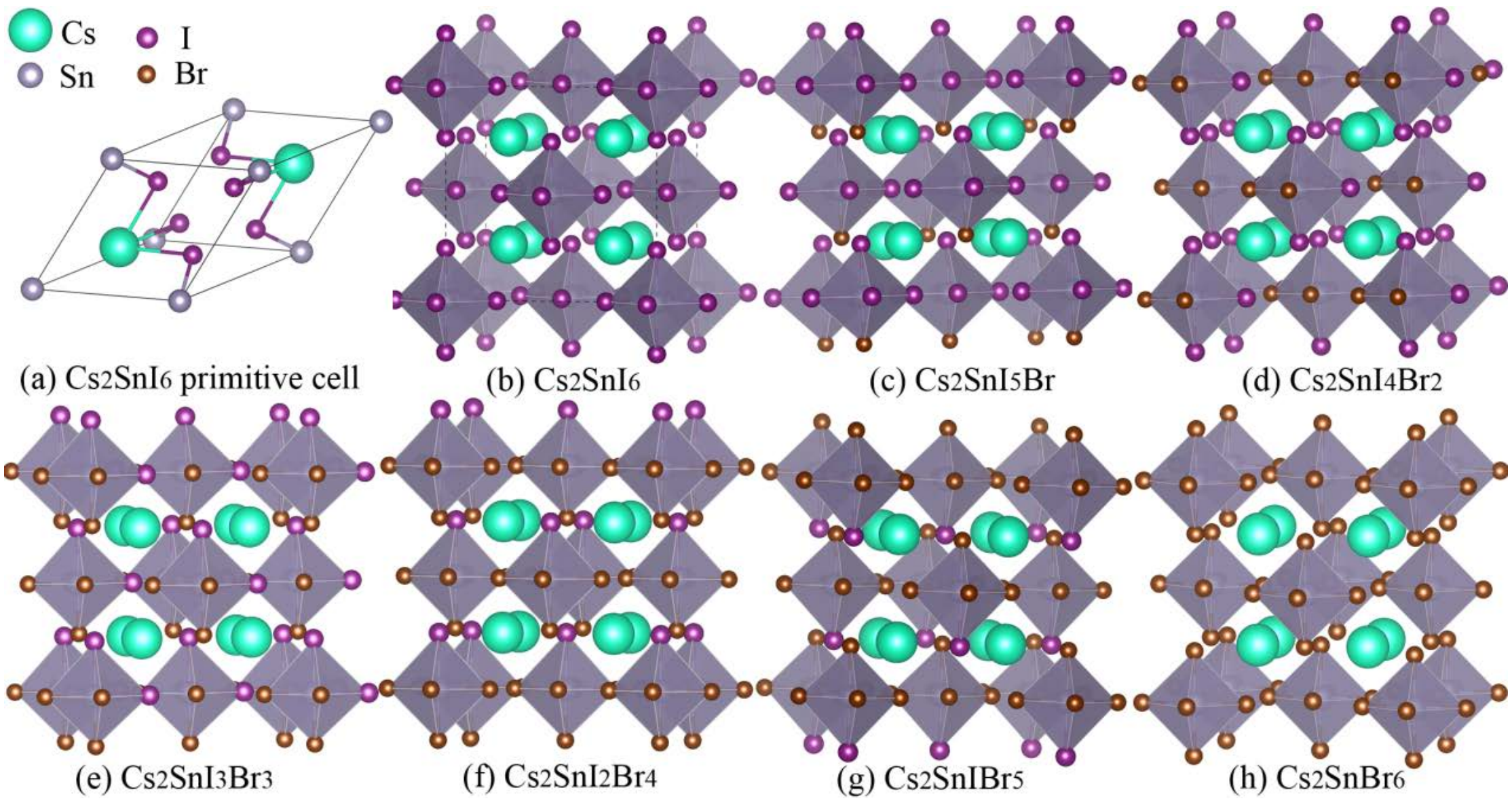

**Fig. 1** Atomic structures of $Cs_2SnI_6$ primitive cell (a) and relaxed conventional cells: (b) $Cs_2SnI_6$, (c) $Cs_2SnI_5Br$, (d) $Cs_2SnI_4Br_2$, (e) $Cs_2SnI_3Br_3$, (f) $Cs_2SnI_2Br_4$, (g) $Cs_2SnIBr_5$, (h) $Cs_2SnBr_6$.

Since the ionic radii and Pauling electronegativity of $F^-$, $Cl^-$ and $Br^-$ ions are different with that of $I^-$ ion, the atomic and electronic structures of $Cs_2SnI_{6-n}X_n$ (X=F, Cl, Br; n= 1~6) should be also different with that of $Cs_2SnI_6$. In the same six coordination number (CN), the Shannon ionic radii of $F^-$, $Cl^-$, $Br^-$ and $I^-$ ion is 1.33Å, 1.81Å, 1.96Å and 2.20Å [30], respectively, while their electronegativity is 3.98, 3.16, 2.96 and 2.66 [31], respectively. These differences will make their Sn-X and Cs-X bond length to be different with each other and the crystal structures will distort accordingly. We measure the lattice constants, the bond lengths of Sn-X and Cs-X basing on the relaxed $Cs_2SnI_{6-n}X_n$ (X=F, Cl, Br; n= 0~6) conventional cells (CIF files in Supplemental Information) and draw them as a function of *n*, as shown in Fig. 2. For the $Cs_2SnI_{6-n}X_n$ (X=F, Cl, Br; n=1~5) compounds, their lattice constants in *xyz* directions and the bond lengths of six Sn-X bonds and twelve Cs-X bonds are always different with each other. In this case, we choose their mean values and the standard deviation is also indicated by the error bar.

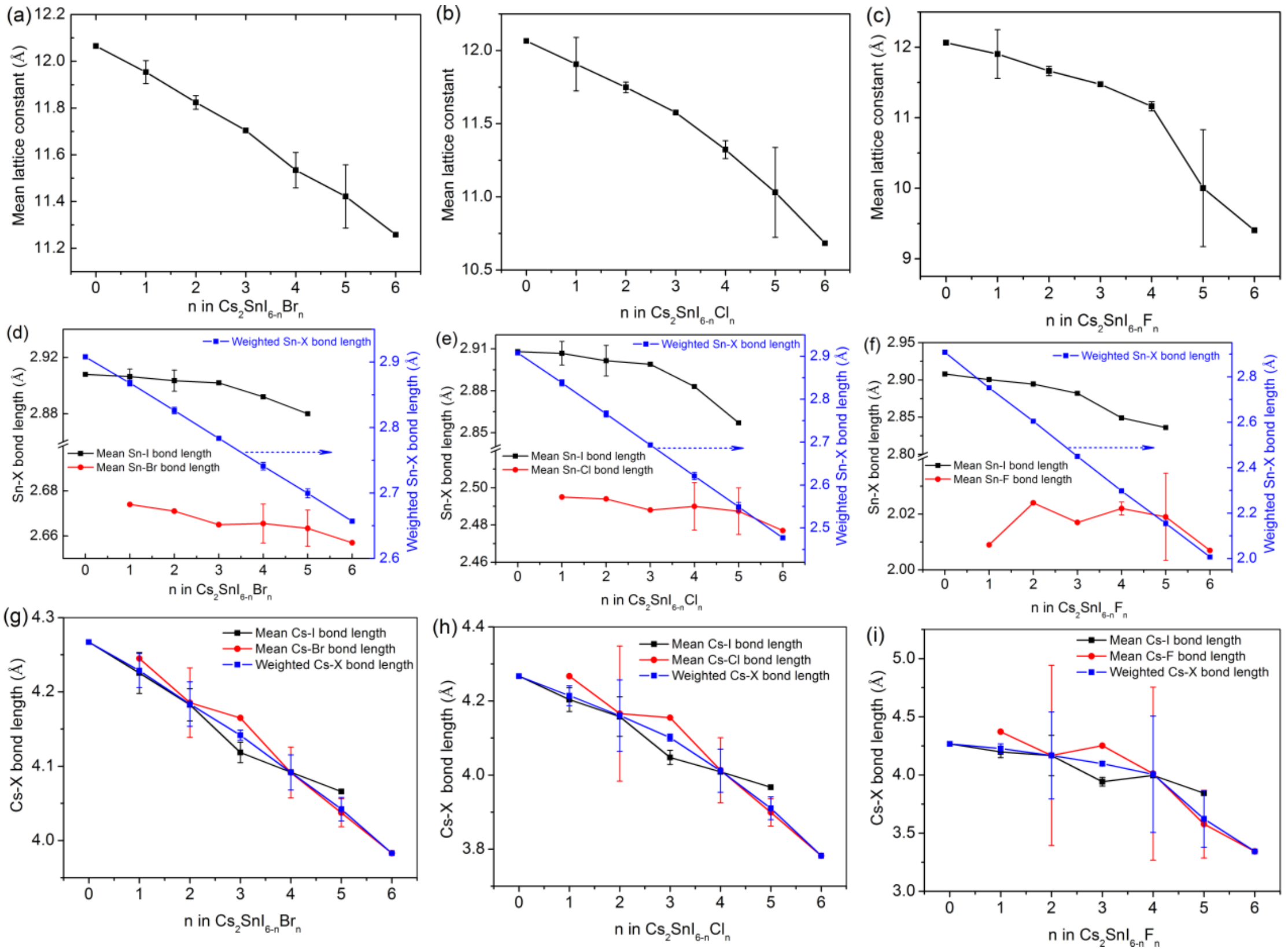

**Fig. 2** Atomic structure parameters of relaxed $Cs_2SnI_{6-n}X_n$. (a)-(c) mean lattice constants; (d)-(f) mean Sn-I, Sn-Br (or Sn-Cl, or Sn-F) and weighted Sn-X bond lengths; (g)-(i) mean Cs-I, Cs-Br (or Cs-Cl, or Cs-F) and weighted Cs-X bond lengths.

As can be seen from Fig. 2(a)-(c) that, the mean lattice constants of $Cs_2SnI_{6-n}X_n$ (X=F, Cl, Br; n=0~6) compounds decrease with the increase of alloying level *n*. This can be easily understood: when $I^-$ ion is substituted by the smaller size of $F^-$, or $Cl^-$, or $Br^-$ ion, the lattice constant will shrink accordingly; the more the $F^-$, or $Cl^-$, or $Br^-$ ion substitutes, the smaller the mean lattice constant is. Karim et al. [14], Lee et al. [17], Yuan et al. [18] and Yang et al. [19] all found that the lattice constant of $Cs_2SnI_{6-x}Br_x$ decrease linearly as Br increases, while Zhu et al. [12] reported that the lattice constant of $Cs_2SnI_xCl_{6-x}$ alloy increases linearly with I/(I+Cl) ratio (<0.2). Such experimental variations of lattice constants of $Cs_2SnI_{6-x}Br_x$ and $Cs_2SnI_xCl_{6-x}$ are consistent with our theoretically calculated results. For $Cs_2SnI_{6-n}F_n$ compound, however, there is still no experimental data available now. Moreover, we find in Fig. 2(c) that its lattice constant decreases linearly when *n* is less than 4, but drops suddenly when *n* is 5 and 6, which situation is not happening for $Cs_2SnI_{6-n}Br_n$ and $Cs_2SnI_{6-n}Cl_n$. This can be attributed to the much larger difference of bond length between Sn-F (Cs-F) and Sn-I (Cs-I) than that between Sn-Br (Cs-Br) or Sn-Cl (Cs-Cl) and Sn-I (Cs-I). Their bond lengths are summarized in Table S1. The difference of bond length between Sn-I and Sn-Br, or Sn-Cl, or Sn-F is -0.251 Å (-8.63%), -0.431 Å (-14.82%), -0.901 Å (-30.98%), respectively. The difference of bond length between Cs-I and Cs-Br, or Cs-Cl, or Cs-F is -0.284 Å (-6.66%), -0.485 Å (-11.37%), -0.924 Å(-21.65%), respectively.

The $Sn^{4+}$ in $Cs_2SnI_6$ bonds with six $I^-$ ions in an octahedra structure. After relaxing the alloyed $Cs_2SnI_{6-n}X_n$ (X=F, Cl, Br; n=0~6) structure completely, the bond length of Sn-I, Sn-Br (or Sn-Cl, or Sn-F) and the weighted Sn-X are measured and shown in Fig. 2(d)-2(f). As the alloying level *n* increases, the weighted Sn-X bond length decreases linearly from 2.908Å for Sn-I to 2.657Å for Sn-Br (Fig. 2(d)), or to 2.477Å for Sn-Cl (Fig. 2(e)), or to 2.007Å for Sn-F (Fig. 2(f)). Since $X^-$ ion is coordinated with the same cation $Sn^{4+}$, the Sn-X bond length will depend on the Shannon ionic radii of $X^-$ ion and so it also decreases in the order: Sn-I> Sn-Br>Sn-Cl>Sn-F. The more the Sn-X is formed, the smaller the weighted Sn-X bond length is. After the Rietveld refinements of XRD data for $Cs_2SnI_{6-n}Br_n$ alloy, Karim et al. [14], Yuan et al. [18], and Yang et al. [19] all found that Sn-X bond length decreases with the increase of $Br^-$ alloying level, which trend is agree well with that of the weighted Sn-X bond length (Fig. 2(d)). For the $Cs_2SnI_{6-n}Cl_n$ and $Cs_2SnI_{6-n}F_n$ alloys, there is no refined Sn-Cl or Sn-F bond length reported till now. On the other hand, we find that the bond length of each Sn-I, Sn-Br and Sn-Cl also decreases slowly: Sn-I bond length decreases from 2.908Å in $Cs_2SnI_6$ to 2.88Å (-0.96%) in $Cs_2SnIBr_5$, or to 2.857Å (-1.75%) in $Cs_2SnICl_5$, or to 2.836Å (-2.48%) in $Cs_2SnIF_5$, respectively; Sn-Br bond length decreases from 2.674Å in $Cs_2SnI_5Br$ to 2.657Å (-0.63%) in $Cs_2SnBr_6$; Sn-Cl bond length decreases from 2.495Å in $Cs_2SnI_5Cl$ to 2.477Å (-0.72%) in $Cs_2SnCl_6$. Sn-X bond length is determined by the sum of ion radii of $Sn^{4+}$ and $X^-$. When the $I^-$ ion is replaced by $Br^-$ (or $Cl^-$, or $F^-$) ion, the covalent radius of $S^{4+}$ ion decreases slowly due to the larger electronegativity and thus the stronger ability to capture valence electron of $S^{4+}$ ion. The more the latter ion is coordinated, the smaller radius the $S^{4+}$ ion has; the larger electronegativity the $X^-$ ion has, the shorter bond length the Sn-X (for n=6) and Sn-I (for n=5) have, as shown in Fig. 2(d)-2(f). Following this hypothesis, the Sn-F bond length for n=1 should be longer than that for n=2. But the former is 2.009 Å, which is shorter than that of the latter (2.024 Å). This is only one case that does not follow our hypothesis above. The reason is still unclear and needs the further study. Yang et al. [19] performed the Raman measurement on $Cs_2SnI_{6-X}Br_X$ (x=0-6) materials and observed the slight shift of Sn-I and Sn-Br Raman peaks (Fig. 5 in Ref. 19). According to the harmonic oscillator model, the Raman shift corresponds to the distance change between two atoms [19]. We find that the shifting trend of the Raman peaks agrees well with the variations of Sn-I and Sn-Br bond length shown in Fig. 2(d).

$Cs^+$ ion in $Cs_2SnI_6$ locates inside the cuboctahedral interstices and bonds with twelve $I^-$ ions. After $I^-$ ions are replaced by $Br^-$ (or $Cl^-$, or $F^-$) ion with smaller ion radius, the volumes of Sn-X octahedral and their cuboctahedral interstices will both become smaller. As a result, the distance between $Cs^+$ ion and surrounding $X^-$ ion, namely Cs-X bond length, also decreases, which is demonstrated in Fig. 2(g)-2(i). Cs-I bond length decreases from 4.267 Å in $Cs_2SnI_6$ to 4.066 Å (-4.71%) in $Cs_2SnIBr_5$ alloy, or to 3.967 Å (-7%) in $Cs_2SnICl_5$ alloy, and or to 3.843 Å (-9.94%) in $Cs_2SnIF_5$ alloy. Cs-Br bond length decreases from 4.245 Å in $Cs_2SnI_5Br$ to 3.983 Å (-6.2%) in $Cs_2SnBr_6$, Cs-Cl bond length decreases from 4.267 Å in $Cs_2SnI_5Cl$ to 3.782 Å (-11.4%) in $Cs_2SnCl_6$, and Cs-F bond length decreases from 4.371 Å in $Cs_2SnI_5F$ to 3.343 Å (-23.5%) in $Cs_2SnF_6$. The weighted Cs-X bond length also decreases from 4.267 Å for Cs-I bond length in $Cs_2SnI_6$ to 3.983 Å (-6.66%) for Cs-Br bond length in $Cs_2SnBr_6$ alloy, or to 3.782 Å (-11.37%) for Cs-Cl bond length in $Cs_2SnCl_6$ alloy, and or to 3.343 Å (-21.65%) for Cs-F bond length in $Cs_2SnF_6$ alloy. Yang et al. [19] reported that Cs-X bond length in $Cs_2SnI_{6-x}Br_x$ (x= 0-6) decreases linearly from ~4.13 Å to ~3.91 Å (-5.33%), which is qualitatively consistent with that shown in Fig. 2(g).

### 3.2. *Electronic structures and optical absorption of $Cs_2SnI_{6-n}X_n$ alloys*

#### *3.2.1. The $Cs_2SnI_{6-n}Br_n$ alloys*

After the alloying of $Br^-$ ions into the $Cs_2SnI_6$, its electronic structure changes accordingly. Fig. 3(a)-3(g) show the total density of states (TDOS) of $Cs_2SnI_{6-n}Br_n$ (n=0~6). We set the valence band maximum (VBM) as Fermi level ($E_f$). Bandgap energy ($E_g$) is defined as the energy difference between VBM and conduction band minimum (CBM). $E_g$ is also labeled inside Fig. 3(a)-3(g) and summarized in Fig. 3(h). The calculated bandgap of $Cs_2SnI_6$ is 1.05 eV, which is close to the experimental value of ~1.3 eV [10-14]. Fig. 3(h) shows that $E_g$ increases nonlinearly with the number of the alloyed $Br^-$ ions, which trend has also been found by several other authors in experiments [14, 17-19].

Since the $E_g$ of $Cs_2SnI_{6-n}Br_n$ (n=1~5) are all between 1.0 eV and 2.0 eV, they can be potentially used as the light absorption materials for inorganic perovskite solar cells and should be studied further. Partial DOS (PDOS) and optical absorption spectra of $Cs_2SnI_{6-n}Br_n$ (n=0~5) are also calculated, as show in Fig. 4 and 5, respectively. For the $Cs_2SnI_6$, CBM mainly consists of I 5p and Sn 5s orbitals and VBM comes mainly from the I 5p orbital. When one $I^-$ ion is replaced by $Br^-$ ion, as shown in Fig. 4(b), Br 4p orbital begins to produce two strong PDOS peaks at ~ -1.0 eV and ~ -2.0 eV inside VB, and a weak PDOS peak at ~2.1 eV inside CB. In this alloying level, Br 4p orbital only have a little contribution to VBM, but a noticeable contribution to CBM. In CB, its PDOS has a higher onset energy than those of I 5p and Sn 5s. As a result, $Cs_2SnI_5Br$ has a larger $E_g$ and a blue-shift absorption edge in comparison with that of $Cs_2SnI_6$ (Fig. 5(b)). When more and more $I^-$ ions are replaced by $Br^-$ ions, PDOS of I 5p orbital is substituted gradually by that of Br 4p orbital, as shown in Fig. 4(c)-4(f). At the meantime, $E_g$ increases from 1.3 eV for $Cs_2SnI_4Br_2$ to 1.99 eV for $Cs_2SnIBr_5$. On the other hand, we find that Sn 5s, I 5p or Br 4p have the same energy range in CB or VB, and they contribute mainly to CBM or VBM. So we consider that CBM and VBM come mainly from the hybrid orbitals of Sn-I(Br) octahedral structure, very little from the Cs-I cuboctahedral interstices structures. There are two absorption peaks appearing in Fig. 5(c)-5(f), and they shift with the number of alloyed $Br^-$ ions. Basing on the analysis of Fig. 4(c)-4(f), we consider that they come from the electron transition from the VBM and PDOS peak at ~ -1.0eV to the CB (~2.5 eV), respectively. The blue shifting of absorption peak is due to the upward moving of CBM when more and more $Br^-$ ions are alloyed.

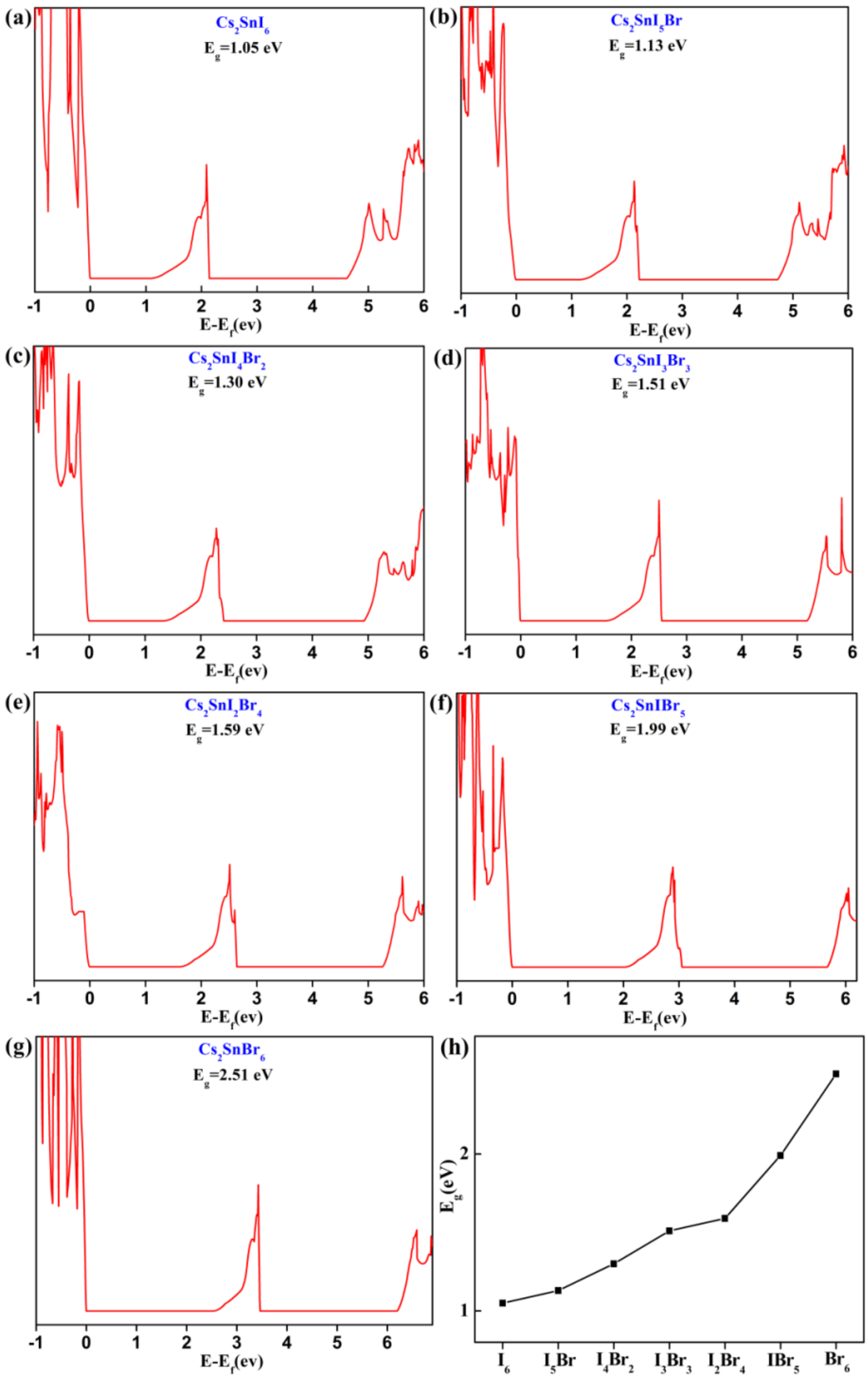

Fig. 3 (a)-(g) TDOS images of $Cs_2SnI_6$ with different numbers of alloyed $Br^-$ ions. $E_g$ is also labeled inside each figure; (h) $E_g$ is shown as a function of the number of the alloyed $Br^-$ ions.

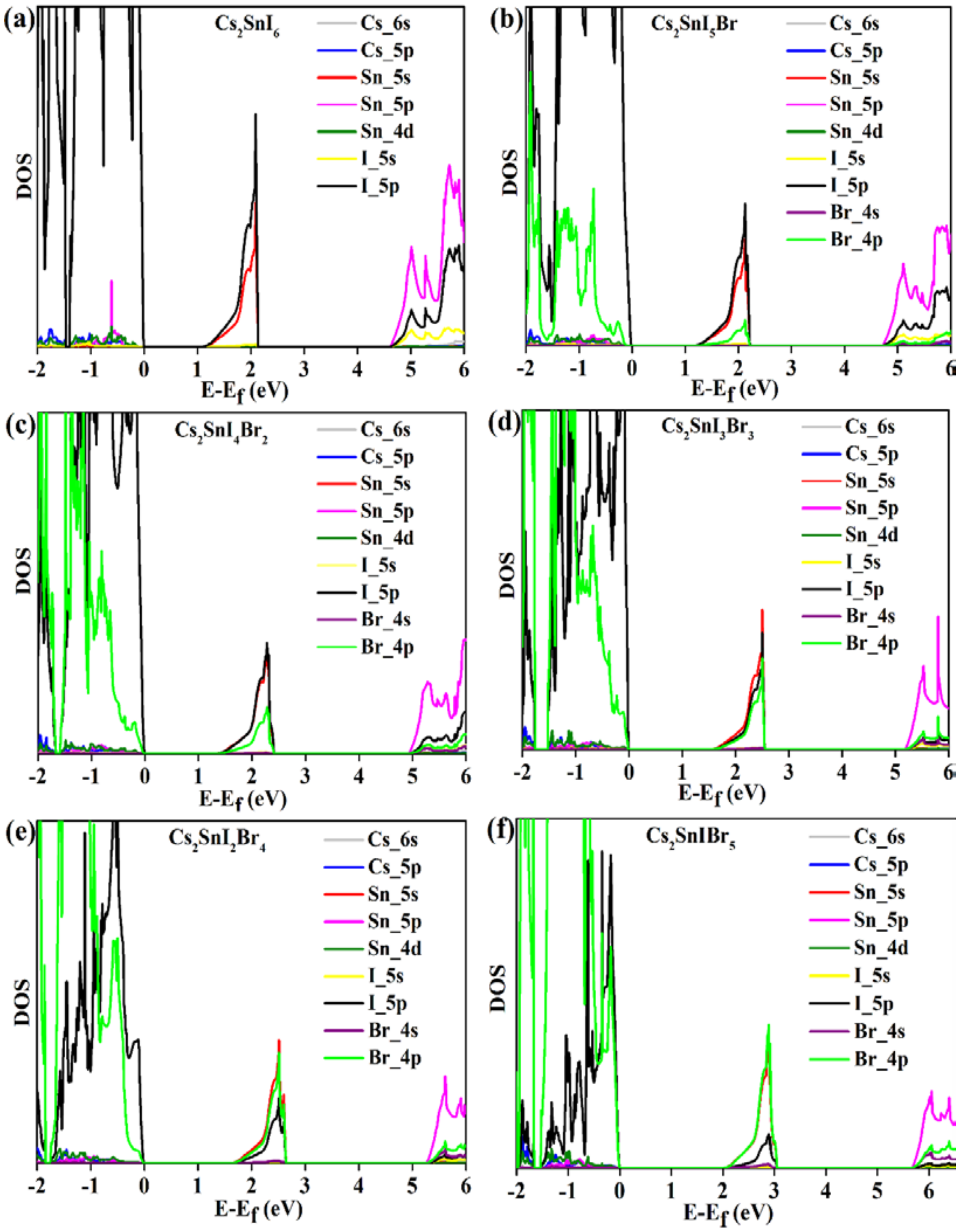

Fig. 4 PDOS images of $Cs_2SnI_6$ (a); $Cs_2SnI_5Br$ (b); $Cs_2SnI_4Br_2$ (c); $Cs_2SnI_3Br_3$ (d); $Cs_2SnI_2Br_4$ (e); $Cs_2SnIBr_5$ (f).

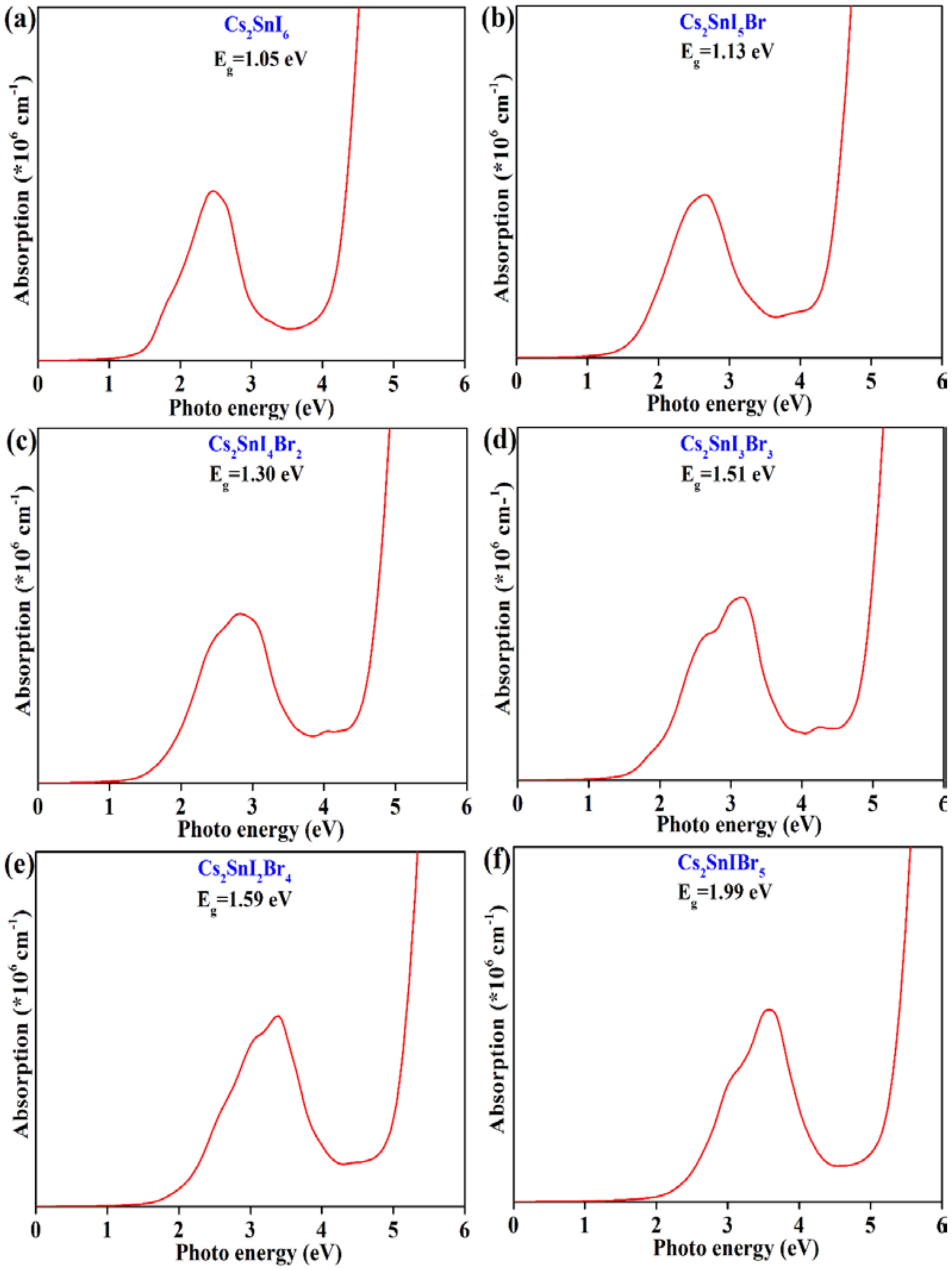

Fig. 5 Calculated light absorption spectra of $Cs_2SnI_6$ (a), $Cs_2SnI_5Br$ (b), $Cs_2SnI_4Br_2$ (c), $Cs_2SnI_3Br_3$ (d), and $Cs_2SnI_2Br_4$ (e); $Cs_2SnIBr_5$ (f).

*3.2.2 The $Cs_2SnI_{6-n}Cl_n$ (n=0~6) alloys*

The TDOS images of $Cs_2SnI_{6-n}Cl_n$ (n=0~6) are shown in Fig. 6(a)-(g). Fig. 6(h) shows the variation of $E_g$ with the number of alloyed $Cl^-$ ion. It can be seen from Fig. 6(h) that $E_g$ increases nonlinearly with the alloyed $Cl^-$ ions, similar with that case for $Br^-$ alloying. The nonlinear increase of $E_g$ with the Cl contents in the Cl-riched $Cs_2SnI_{6-n}Cl_n$ has also been observed by Zhu et al. [12] in experiments.

As shown in Fig. 7, PDOS of $Cs_2SnI_5Cl$, $Cs_2SnI_4Cl_2$, $Cs_2SnI_3Cl_3$ and $Cs_2SnI_2Cl_4$ are also calculated because that their $E_g$ are between 1.0 eV and 2.0 eV and can potentially be used as the light absorber for perovskite solar cells. After replacing one $I^-$ ion in $Cs_2SnI_6$ by $Cl^-$ ion, as shown in Fig. 7(b), Cl 3p orbital generates a strong PDOS peak at the energy range of -0.5~ -1.5 eV and a weak PDOS peak at ~ 2.2 eV. The low energy branch of Cl 3p is the bonding orbital while the high energy branch is the anti-bonding orbital. The bonding orbital of Cl 3p has a very little contribution to the VBM, which consists mainly of I 5p orbital. The anti-bonding orbital of Cl 3p has a higher onset energy and it contributes partially to the CBM, blue shifting the CBM. This is the reason why $Cs_2SnI_5Cl$ has a larger $E_g$ than that of $Cs_2SnI_6$. When two $I^-$ ions are replaced by $Cl^-$ ions, as shown in Fig. 7(c), the contribution of Cl 3p orbital to the deep VB (-2 eV~ -1.5 eV) is enhanced but its contribution to the VBM is still very little. The anti-bonding Cl 3p has an increasing contribution to CBM while the contribution of I 5p orbital to the CBM decreases at the meantime. When more and more $I^-$ ions are replaced by $Cl^-$ ions, as shown in Fig. 7(d)-7(e), PDOS of I 5p orbital at VBM and CBM are gradually substituted by that of Cl 3p orbital. As a result, the $E_g$ of $Cs_2SnI_3Cl_3$ and $Cs_2SnI_2Cl_4$ enlarge continuously.

The light absorption properties of $Cs_2SnI_5Cl$, $Cs_2SnI_4Cl_2$, $Cs_2SnI_3Cl_3$ and $Cs_2SnI_2Cl_4$ were also calculated, as shown in Fig. 8. Comparing Fig. 8(a) with 8(b), we can find that one vicarious $Cl^-$ ion blue shifts the absorption edge of $Cs_2SnI_6$, consistent with the larger $E_g$ shown in Fig. 6(b). When more and more $I^-$ ions are replaced by $Cl^-$ ions, besides the continuous blue shifting of absorption edge, there are three absorption peak appearing gradually, including one peak at ~2.5eV, one peak at ~3.5 eV, and one peak at ~4.5 eV. By carefully analyzing the PDOS shown in Fig. 7, we consider that the three absorption peaks originate from the electron transition from the VBM, the PDOS peak at ~ -1.0 eV and ~ -2.0 eV, respectively, to the DOS peak at ~2.5 eV in CB.

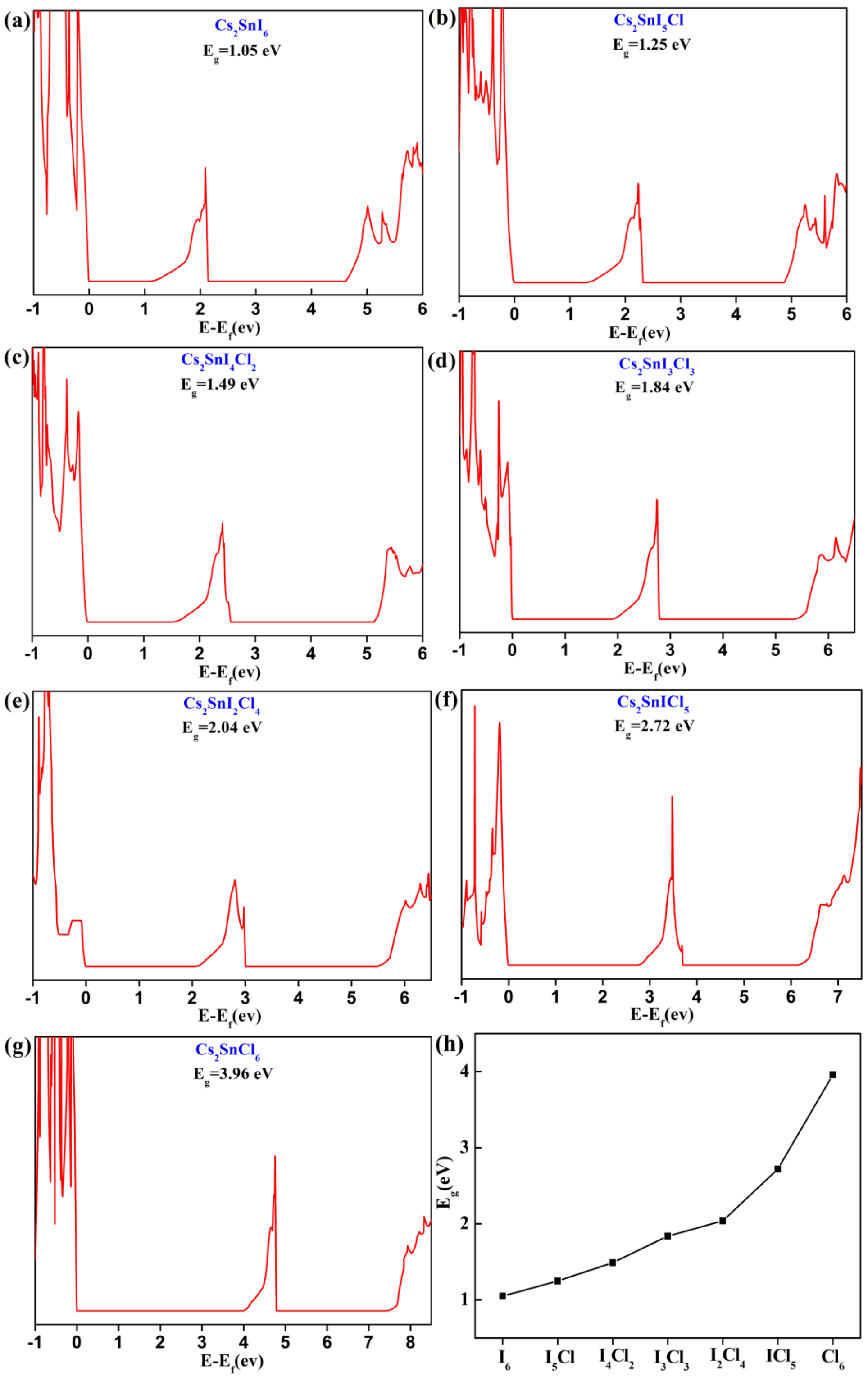

Fig. 6 (a)-(g) TDOS images of $Cs_2SnI_6$ with different numbers of alloyed $Cl^-$ ions. $E_g$ is also labeled inside each figure; (h) $E_g$ is shown as a function of the number of alloyed $Cl^-$ ions.

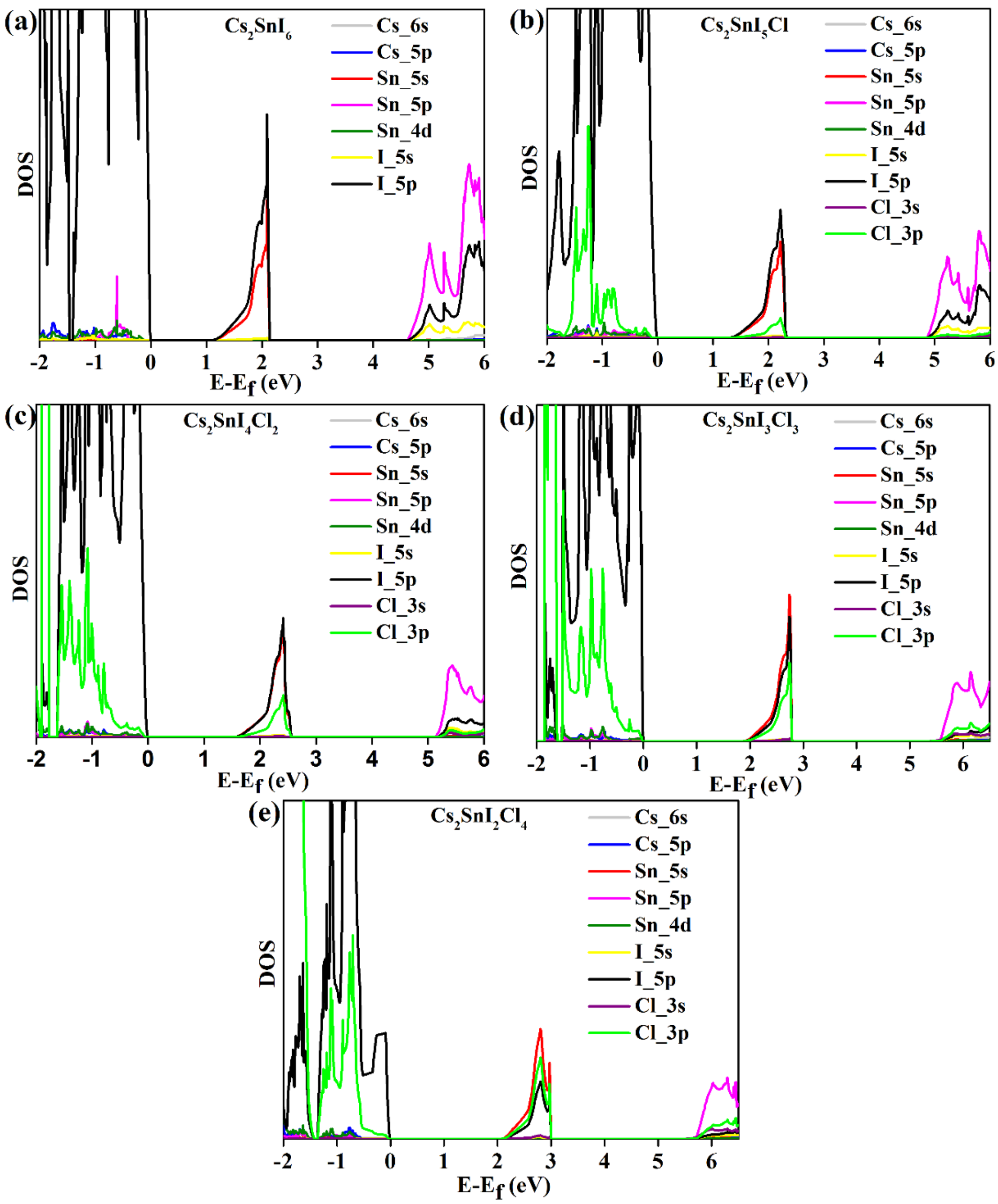

Fig. 7 PDOS images of $Cs_2SnI_6$ (a), $Cs_2SnI_5Cl$ (b), $Cs_2SnI_4Cl_2$ (c), $Cs_2SnI_3Cl_3$ (d) and $Cs_2SnI_2Cl_4$ (e).

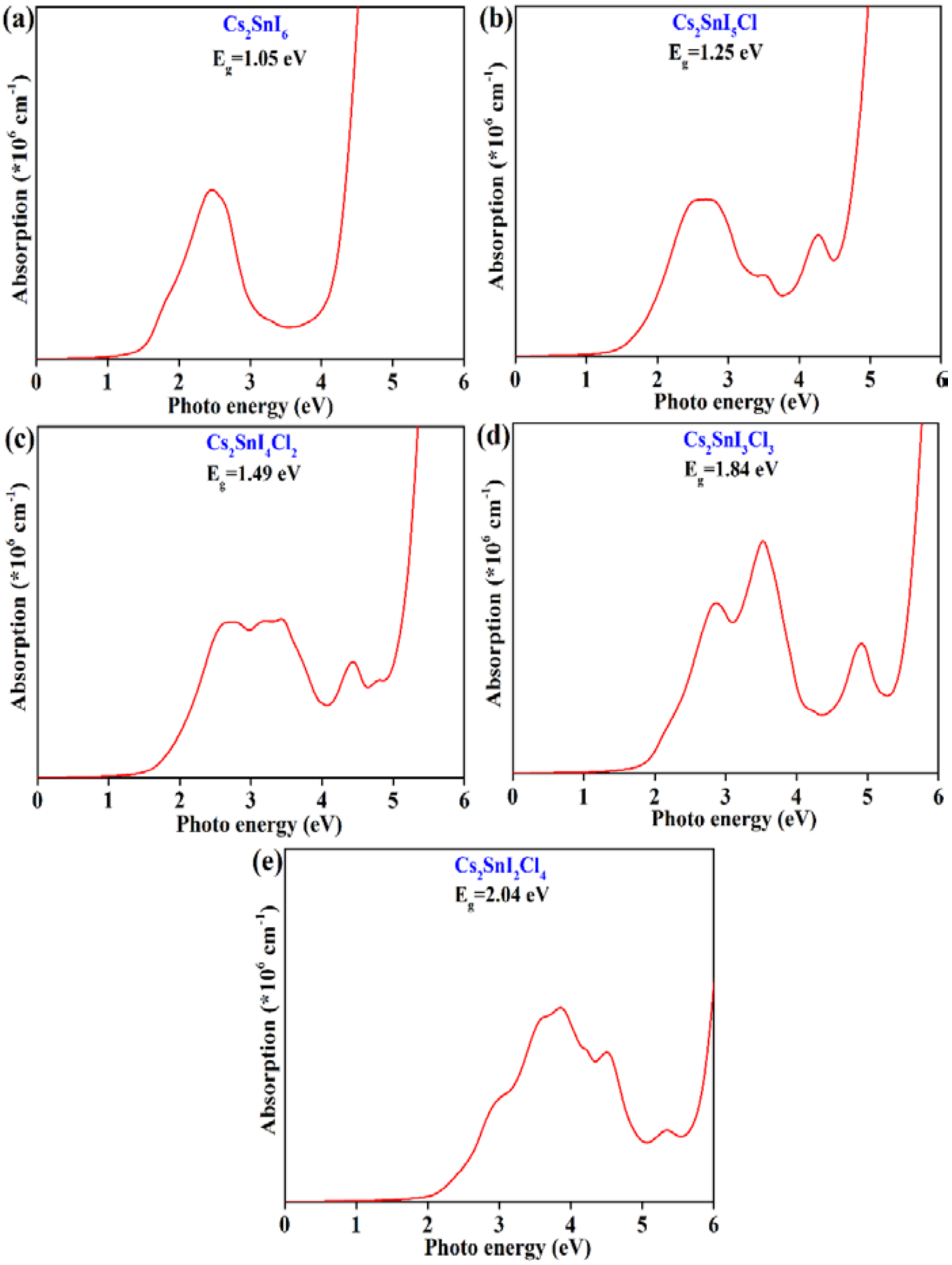

Fig. 8 Calculated light absorption spectra of $Cs_2SnI_6$ (a), $Cs_2SnI_5Cl$ (b), $Cs_2SnI_4Cl_2$ (c), $Cs_2SnI_3Cl_3$ (d), and $Cs_2SnI_2Cl_4$ (e).

### *3.2.3. The $Cs_2SnI_{6-n}F_n$ (n=0~6) alloys*

After the alloying of $F^-$ ions into the $Cs_2SnI_6$, its electronic structure changes remarkably. Fig. 9(a)-9(g) show the TDOS of $Cs_2SnI_{6-n}F_n$ (n=0~6). Fig. 9(h) shows that $E_g$ increases nonlinearly with the alloyed F- ions. Since $E_g$ of $Cs_2SnI_5F$ and $Cs_2SnI_4F_2$ are 1.57 eV and 1.91 eV, respectively, they can potentially be used as the light absorption materials for the inorganic perovskite solar cell. In view of this point, their PDOS are also calculated and shown in Fig. 10. After the alloying of $F^-$ ion, F 2p orbital contributes more and more to the CBM of band structure while the VBM consist mainly of I 5p orbital, as shown in Fig. 10(b) and 10(c). As a result, $E_g$ increase from 1.05 eV for $Cs_2SnI_6$ to 1.57 eV for $Cs_2SnI_5F$ and to 1.91 eV for $Cs_2SnI_4F_2$.

We also calculated the light absorption spectra of $Cs_2SnI_5F$ and $Cs_2SnI_4F_2$, as shown in Fig. 11. After alloying of one $F^-$ ion into $Cs_2SnI_6$, its light absorption edge begins to blue shift, as shown in Fig. 11(a) and 11(b), due to the larger $E_g$ than that of $Cs_2SnI_6$. In addition, a new light absorption peak appears at ~4 eV in Fig. 11(b). According to the careful analysis of the PDOS image shown in Fig. 10(b), we believe that it comes from the electrons transition from the bonding orbital of F 2p (at ~-1.5 eV) to the anti-bonding orbital of F 2p (at ~2.5 eV). When two of $F^-$ ions replace the $I^-$ ions, the light absorption edge continues to blue shift because that the CBM has more contribution from F 2p orbital with the higher energy. At the meantime, the light absorption peak at ~4 eV blue shifts to ~4.5 eV and becomes stronger, which can be ascribed to the electron transition from the stronger PDOS of bonding F 2p orbitals (~ -2.0 eV) to the anti-bonding F 2p orbitals (~ 2.5 eV), as shown in Fig. 10(c).

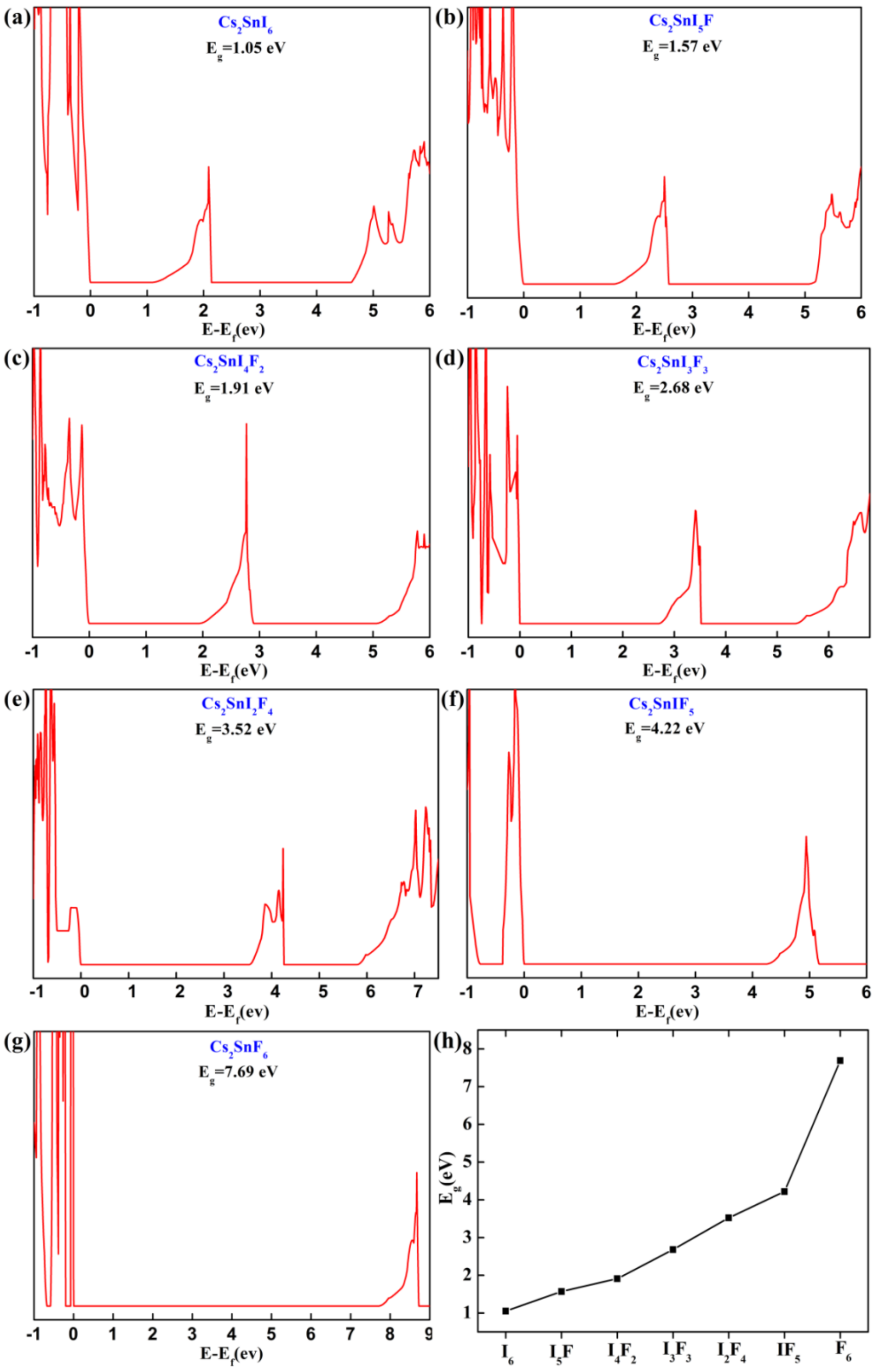

Fig. 9 (a)-(g) TDOS of $Cs_2SnI_6$ with different numbers of alloyed $F^-$ ions. $E_g$ is labeled inside each figure; (h) $E_g$ is shown as a function of the number of alloyed $F^-$ ions.

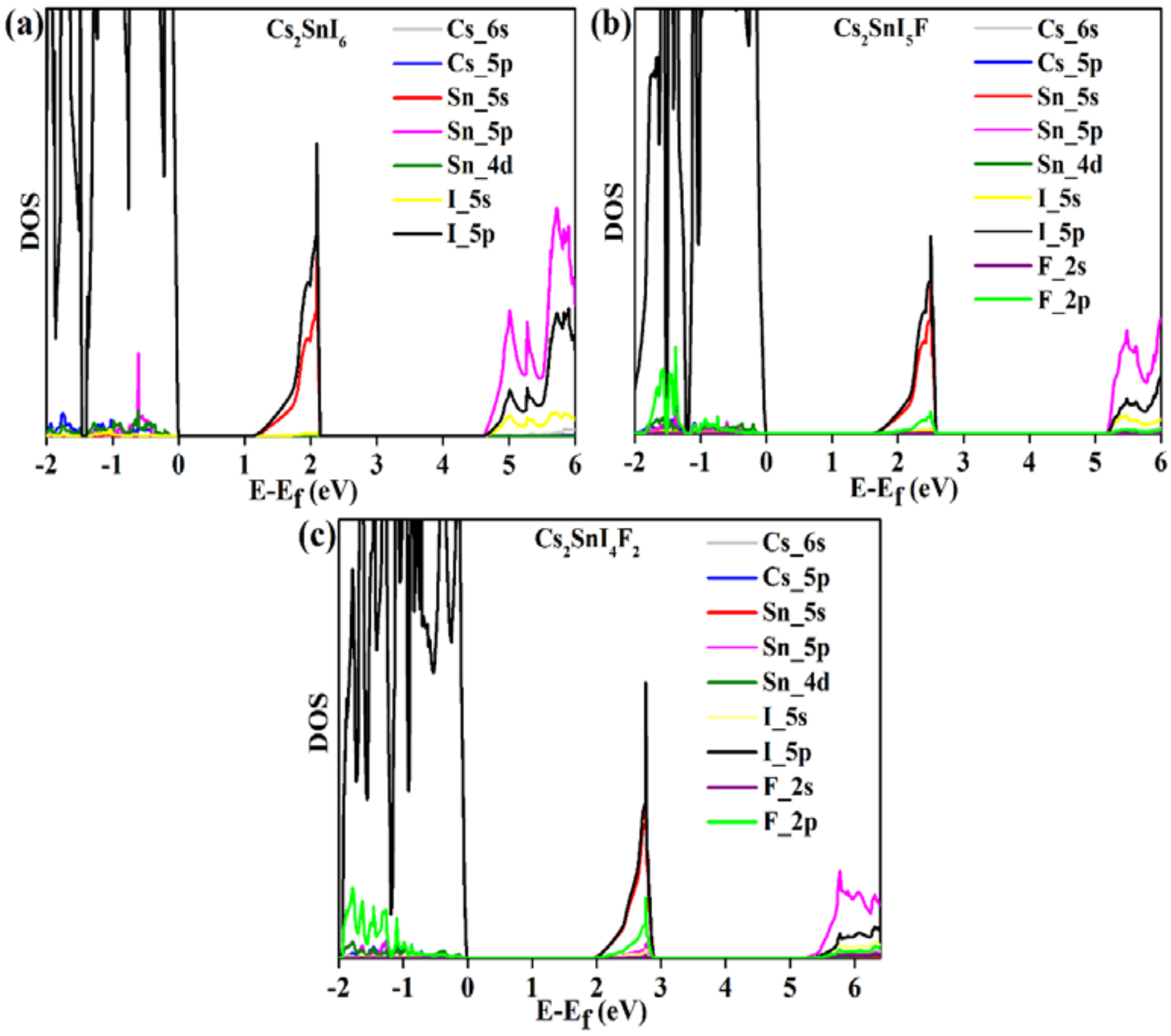

Fig. 10 Partial DOS of $Cs_2SnI_6$ (a), $Cs_2SnI_5F$ (b) and $Cs_2SnI_4F_2$ (c).

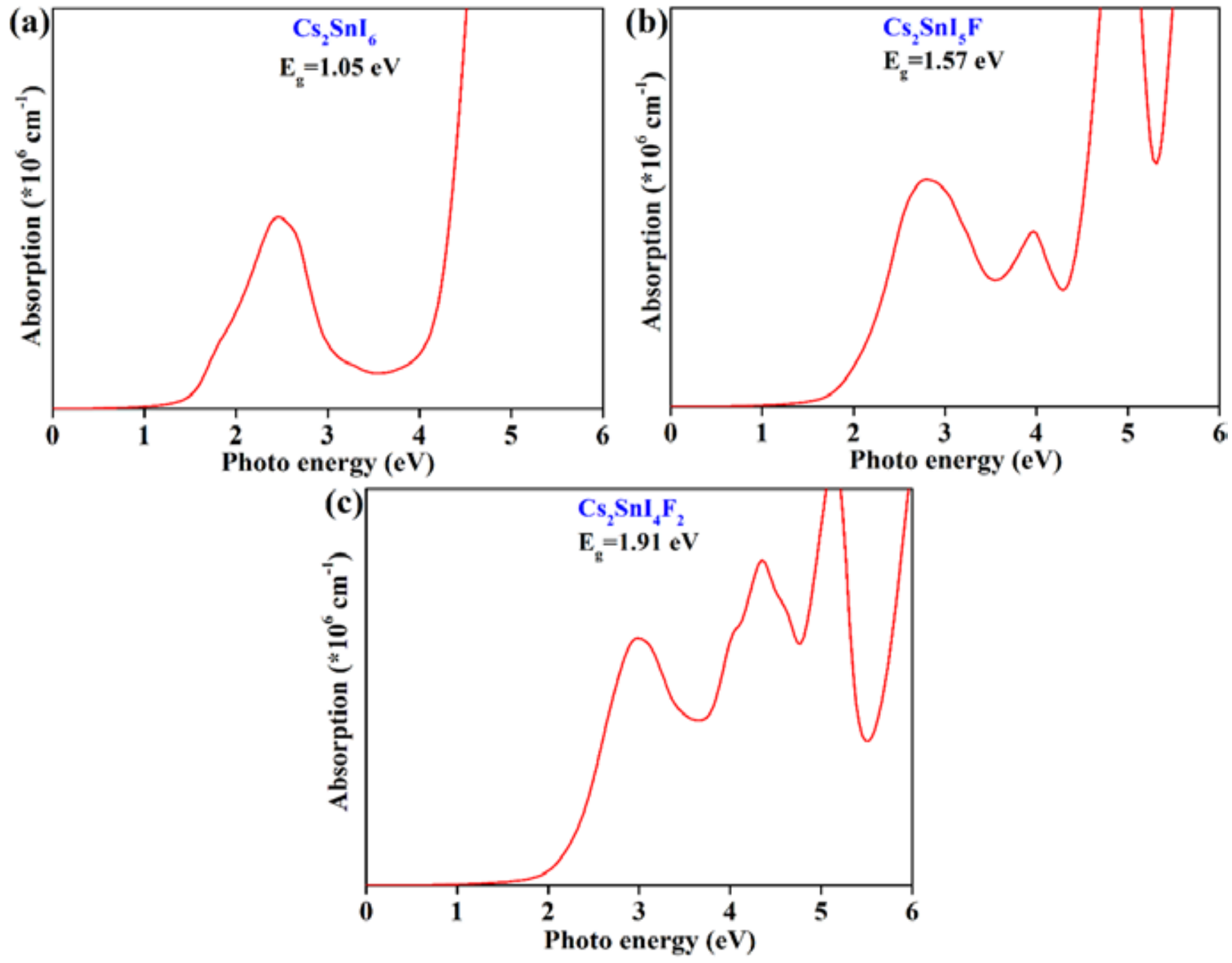

Fig. 11 Calculated light absorption spectra of $Cs_2SnI_6$ (a), $Cs_2SnI_5F$ (b), and $Cs_2SnI_4F_2$ (c).

*3.2.4. The band structures of interesting $Cs_2SnI_{6-n}X_n$ alloys*

Besides the suitable $E_g$ value, it is also important for the eleven interesting $Cs_2SnI_{6-n}X_n$ alloys (1.0 eV<$E_g$<2.0 eV) that whether they have a direct band gap or not. Direct bandgap semiconductors usually have a bigger probability for electron transition between the CBM and VBM than that of indirect bandgap semiconductors, due to the needless participation of phonon. Hence, direct bandgap semiconductors have a larger optical absorption coefficient and thus a thinner film can absorb the enough solar light for solar cells.

Previous many authors have reported that $Cs_2SnI_6$ has a *direct* band gap of 1.011 eV [6], or 0.97 eV [11], or 0.92 eV [14], or 0.93 eV [32] by using the HSE06 hybrid functional. However, it is still unclear that whether the $Cs_2SnI_{6-n}X_n$ alloys are direct bandgap semiconductors or not. So we calculated the band structures of the eleven interesting $Cs_2SnI_{6-n}X_n$ alloys, as shown in Fig. 12 and Fig. S1. In order to find the complete high symmetry *k*-path in the Brillouin zones (BZ), we adopted the methods suggested by Hinuma et al. [33] and Togo et al. [34] and performed at the online website [35]. Fig. 12 and Fig. S1 show that the amount of high symmetry k points in BZ increases largely due to the lattice distortion and the reduced symmetry after the alloying of other halide ions into $Cs_2SnI_6$. Interestingly, the eleven $Cs_2SnI_{6-n}X_n$ alloys all have a *direct* bandgap. We also calculated the orbital projected band structure in order to clarify the contribution of each atomic orbital to each band in $Cs_2SnI_{6-n}X_n$ band structure. For $Cs_2SnI_6$, we find that its VBM consists mainly of I 5p orbital and its CBM is constituted by I 5p and Sn 5s orbitals, as shown in Fig. 12(a). Except the $Cs_2SnIBr_5$, the VBM of other $Cs_2SnI_{6-n}X_n$ consists mainly of I 5p orbital. For $Cs_2SnIBr_5$, its VBM is made up of more than half of I 5p and less than half of Br 4p orbital. The CBM of eleven interesting $Cs_2SnI_{6-n}X_n$ all include the contribution from I 5p and Sn 5s orbitals, but the contribution from Br 4p or Cl 3p orbitals increases when the alloying level n≥2.

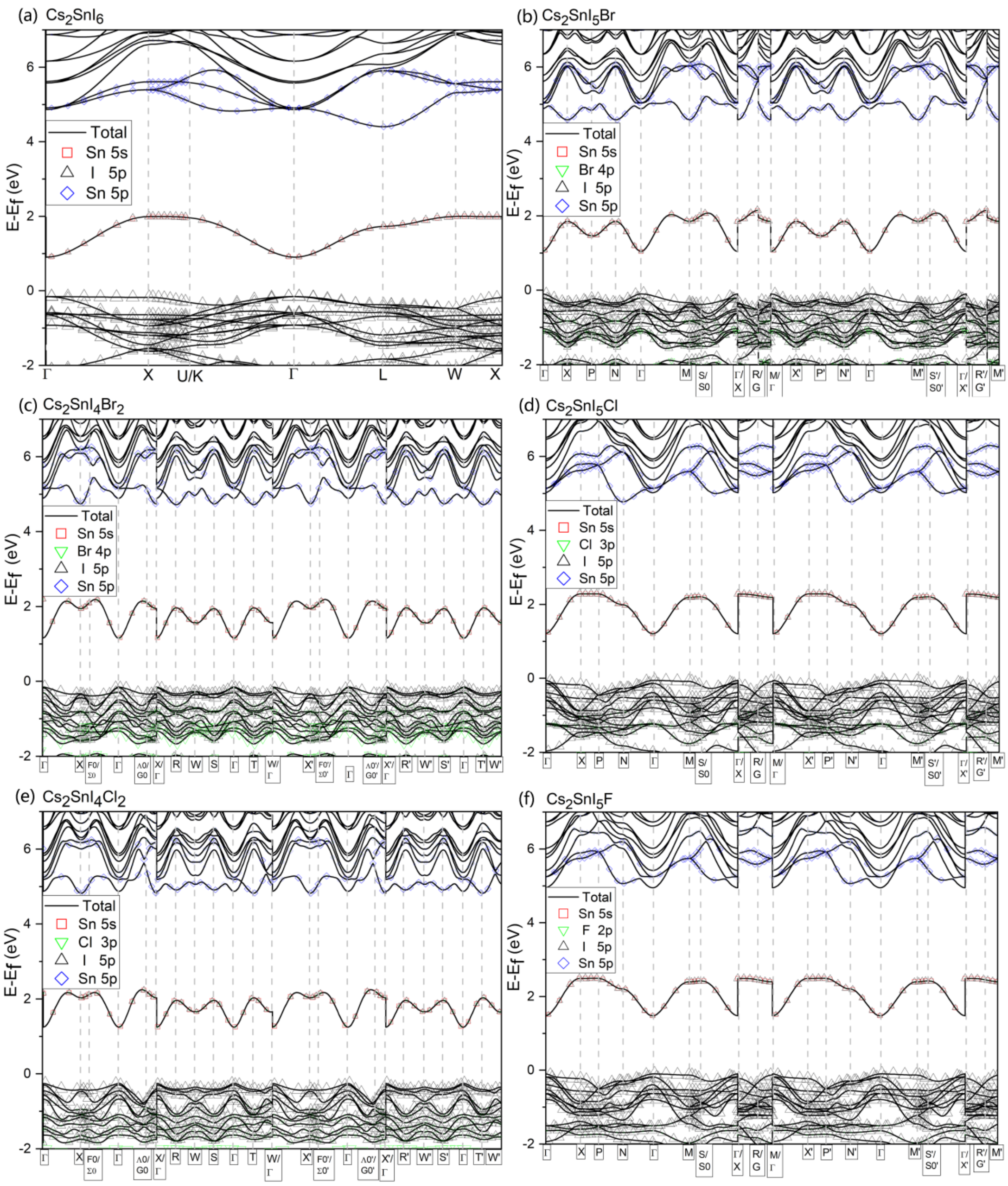

Fig. 12 Electronic band structures of six interesting $Cs_2SnI_{6-n}X_n$ alloys.

Clearly, the electronic band structure of eleven interesting $Cs_2SnI_{6-n}X_n$ alloy is composed of the composition atoms inside the Sn-X octahedral structure. Fig. 1 and supplementary CIF files both show that the Sn-X octahedral atomic structure keeps also unchanged after alloying. In addition, the alloyed halide ions ($Br^-$, or $Cl^-$, or $F^-$) locate at the same *Group VII* with $I^-$ ion in *element periodic table* and thus they should have the similar outmost *p* orbital and bonding properties. This is possibly the reason why the alloying of other halide ions does not transform the $Cs_2SnI_6$ from *direct* to *indirect* bandgap semiconductors.

**4. Conclusions**

In order to find new $Cs_2SnI_6$ derivates for inorganic perovskite solar cells, first-principles calculation on the atomic and electronic structures as well as optical absorption of $Cs_2SnI_{6-n}X_n$ (X=F, Cl, Br; n=0~6) compounds are performed. As the alloying level *n* increases, the mean lattice constants, the weighted Sn-X and Cs-X bond lengths all decreases linearly. In the Sn-X octahedral structure, the bond length of each Sn-X diminishes slightly, possibly due to the reduced covalent radius of $S^{4+}$ ion originating from the stronger ability of $X^-$ (X=F, Cl or Br) ion to draw the outer electron of $S^{4+}$ ion. The $E_g$ of $Cs_2SnI_{6-n}X_n$ increases nonlinearly with the number of the alloyed $X^-$ (X=F, Cl or Br) ions. Eleven $Cs_2SnI_{6-n}X_n$ compounds with $E_g$ between 1.0 eV and 2.0 eV can be potentially used as the light absorption layer for inorganic perovskite solar cells and are further studied. Their PDOS demonstrate that as the alloying level *n* increases, I 5p orbital in VBM and CBM is gradually substituted by Br 4p, or Cl 3p, or F 2p orbital. As a result, their $E_g$ increases consequently and light absorption edge blue shifts continuously. Finally, our calculation shows that the eleven $Cs_2SnI_{6-n}X_n$ alloys all have a *direct* bandgap although the lattice distortion induced by the alloying of $X^-$ ions.

**Declaration of Competing Interest**

The authors declare that they have no known competing financial interests or personal relationships that could have appeared to influence the work reported in this paper.

**Acknowledgments**

This work is supported by Henan Science and Technology Agency (No. 202102210065) and Graduate Education Innovation Plan of Henan Key Laboratory of Photovoltaic Materials (No. CX3040A0950132).

**Appendix A. Supporting information**

Supplementary data associated with this article can be found in the online version.